\documentclass[reprint, amsmath, amssymb, prb, cha, superscriptaddress]{revtex4-2}

\usepackage{graphicx}
\usepackage{dcolumn}
\usepackage{bm}
\usepackage{hyperref}
\usepackage{textcomp}
\usepackage{gensymb}
\usepackage{pbox}
\usepackage[all]{nowidow}

\begin{document}

\title{Theoretical insights for Improving the Schottky-barrier Height at the Ga$_2$O$_3$/Pt Interface}

\author{F\'elix Therrien}
 \affiliation{Colorado School of Mines}
\affiliation{National Renewable Energy Laboratory}
\author{Andriy Zakutayev}%
\affiliation{National Renewable Energy Laboratory}
\affiliation{Colorado School of Mines}%
\author{Vladan Stevanovi\'c}%
 \email{vstevano@mines.edu}
\affiliation{Colorado School of Mines}%
\affiliation{National Renewable Energy Laboratory}

\date{\today}

\begin{abstract}
In this work we study the Schottky barrier height (SBH) at the junction between $\beta$-Ga$_2$O$_3$ and platinum, a system of great importance for the next generation of high-power and high-temperature electronic devices. Specifically, we obtain interfacial atomic structures at different orientations using our structure matching algorithm and compute their SBH using electronic structure calculations based on hybrid density functional theory.  The orientation and strain of platinum are found to have little impact on the barrier height. In contrast, we find that decomposed water (H.OH), which could be present at the interface from Ga$_2$O$_3$ substrate preparation, has a strong influence on the SBH, in particular in the ($\overline{2}$01) orientation. The SBH can range from $\sim$2 eV for the pristine interface to nearly zero for the full H.OH coverage. This result suggests that SBH of $\sim$2~eV can be achieved for the Ga$_2$O$_3$($\overline{2}$01)/Pt junction using the substrate preparation methods that can reduce the amount of adsorbed water at the interface.
\end{abstract}

\maketitle

\section{Introduction}

It is estimated that, currently, about 30\% of all electricity in the United States goes through high-power electronic devices such as AC-DC and DC-AC converters used in battery chargers, power supplies, etc. \cite{tolbert2005power} This proportion could rise up to 80\% in the next decade \cite{heuberger2018real} given the expected proliferation of renewable energy technologies, modernization of the power grids and transition to electric transportation. Therefore, even marginal gains in power device efficiency could have a major impact on the aggregate energy consumption. 

Wide bandgap semiconductors such as SiC, GaN and Ga$_2$O$_3$ with their higher breakdown fields and good conductivity have emerged as alternatives to standard inexpensive, but difficult to engineer Si-based power devices \cite{gorai2019computational}. In particular, $\beta$-Ga$_2$O$_3$ has been identified as the fourth generation of high-power electronic materials because of its relatively low projected cost \cite{reese2019much}, wide bandgap (4.8 eV) \cite{peelaers2015brillouin, orita2000deep}, large breakdown field and high permittivity which together gives it one of the highest Baliga figure of merit \cite{higashiwaki2012gallium, higashiwaki2016recent}, the performance indicator for high-power electronics. In addition to high temperature and high power electronics, Ga$_2$O$_3$ has applications in gas (humidity) sensing \cite{mazeina2010functionalized, wang2015humidity}, solar-blind photodetection \cite{chen2019review}, (photo)catalysis \cite{jin2015effect, hou2007photocatalytic}, electroluminecense \cite{miyata2000gallium, wellenius2008bright} and many others \cite{pearton2018review, mastro2017perspective}. Because of its unique properties, it has received significant attention from the research community in recent years \cite{pearton2018review}.

Limited doping, however, puts some limitations on the architecture of Ga$_2$O$_3$-based devices. Namely, it is currently possible to dope Ga$_2$O$_3$ n-type, while obtaining p-type doped Ga$_2$O$_3$ is nearly impossible to attain \cite{peelaers2019deep, goyal_2021:arxiv}. In fact, wide bandgap and bipolar materials are generally very hard to come by \cite{goyal_CM:2020}. Consequently, device prototyping has focused mainly on (unipolar) Schottky barrier diodes (SBD) and metal-oxide-semiconductor field-effect transistors (MOSFETs) \cite{xue2018overview}. Both rely on the formation of an electric charge depletion region, a Schottky barrier, at the interface between a metal and the semiconductor (n-type doped $\beta$-Ga$_2$O$_3$ here). Its height determines the electrical (rectifying) characteristics of the contact. A large Schottky barrier height (SBH) is ideal for diodes which require low current under reverse bias (leakage current). On the other hand, a near-zero SBH where current flows freely through the interface in both directions is ideal for passive (or Ohmic) electrical contacts.

Considering the importance of the SBH for high-power devices, revealing the factors that govern it as well as offering strategies on how to improve it (increase or lower depending on the need) would be invaluable. Specifically, for the $\beta$-Ga$_2$O$_3$/Pt junction studied here, identifying a procedure to obtain large SBH leading to a low leakage current would potentially have lasting impact on the future, Ga$_2$O$_3$-based power devices \cite{harada_APLM:2020, farzana2017influence}.

\begin{figure*}
\includegraphics[width=0.9\linewidth]{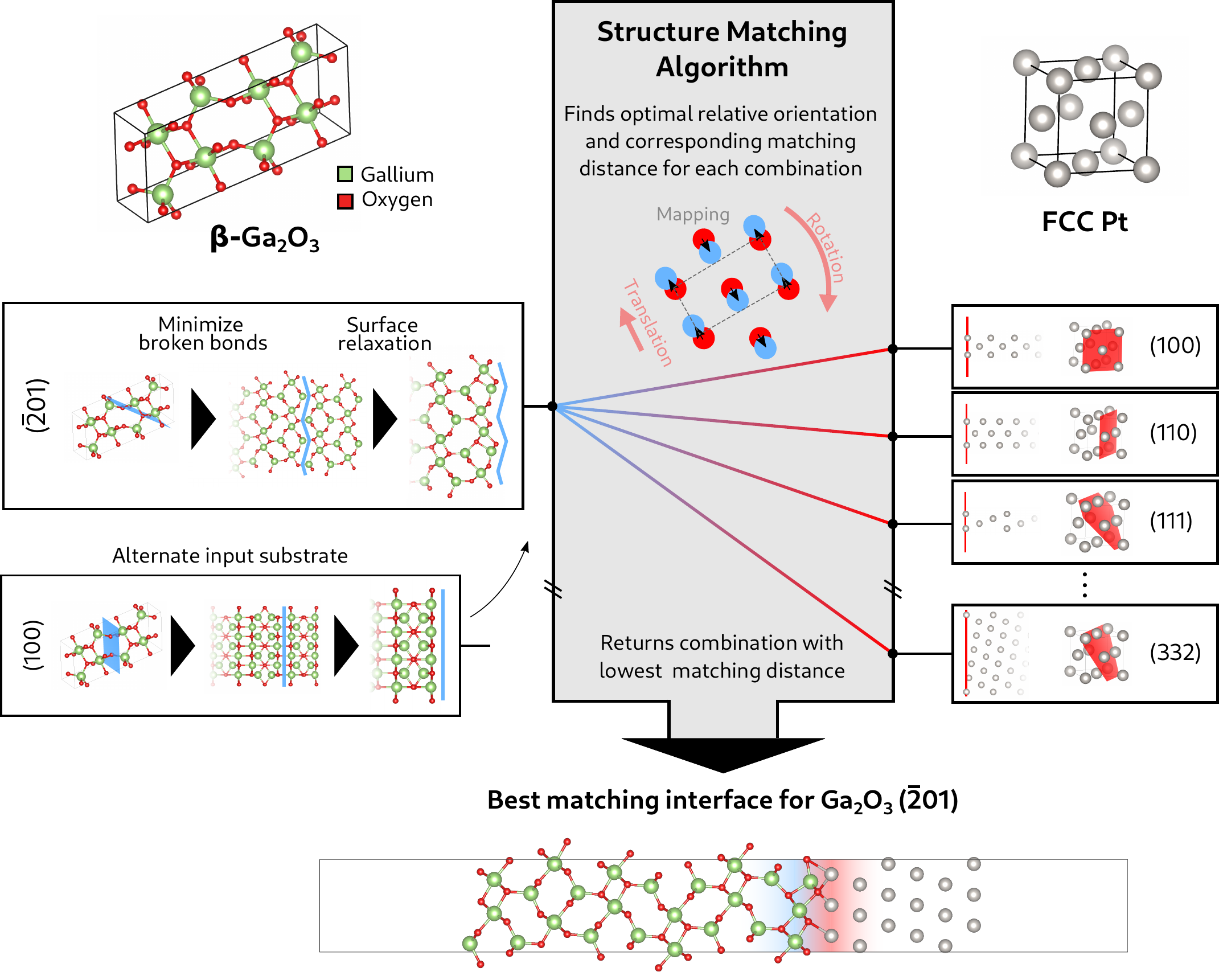}
\caption{\label{fig:workflow}
Workflow to construct interfacial atomic structures. The structure matching algorithm represented by the box in gray takes as an input a $\beta$-Ga$_2$O$_3$ surface (left) and a set of Pt surfaces (right). It returns an interface structure of the best matching of the combinations of interfaces.}
\end{figure*}
In this work, we investigate the Schottky barrier at the $\beta$-Ga$_2$O$_3$/Pt junction using atomistic computational methods. We use our structure matching algorithm\cite{therrien2020matching} to provide the best matching interface structures between platinum and two orientations of gallium oxide, ($\overline{2}$01) and (100). These structures are then used as inputs for electronic calculations at the hybrid density functional theory level, which provide a means to evaluate the barrier heights. 
We find SBHs of $\sim$2~eV for $\beta$-Ga$_2$O$_3$($\overline{2}$01)/Pt and $\sim$1.5~eV for $\beta$-Ga$_2$O$_3$(100)/Pt in comparison with the experimentally measured values of $\sim$1~eV and $\sim$1.4~eV respectively. We then show how the presence of adsorbed dissociated water at the interface can significantly reduce the measured SBH for the $\beta$-Ga$_2$O$_3$($\overline{2}$01)/Pt interface but has little impact on $\beta$-Ga$_2$O$_3$(100)/Pt, which could explain the discrepancy between predicted and measured values in the ($\overline{2}$01) orientation.

Finally, we investigate how the Pt orientation and strain influence the results and find that neither has an appreciable impact on the predicted barrier heights. The results presented in this paper highlight the important role of surface adsorbates on electrical contact properties and provide insight in improving the performance of future Ga$_2$O$_3$ devices.

\section{Methods} \label{sec:methods}

The workflow used in this paper is depicted in Fig.~\ref{fig:workflow}. We started from the crystal structure of $\beta$-Ga$_2$O$_3$ (conventional unit cell with 20 atoms shown in Fig.~\ref{fig:workflow}) and the face-centered cubic (FCC) platinum. Here, we chose platinum because of its high work function which in general leads to higher Schottky barriers. On the $\beta$-Ga$_2$O$_3$ side (left), we first cut the structure along specific orientation that we wished to study. We were particularly interested in the $(\overline{2}01)$ orientation of $\beta$-Ga$_2$O$_3$ because of previous experimental work \cite{tellekamp2020growth, heinselman2021performance}. We also investigated the (100) orientation, the most stable, primarily to validate our approach against previous studies \cite{xu2020first}. Both of these orientations have multiple possible surface terminations. We chose the terminations that minimize the number of broken bonds, where bonds are defined geometrically as those within the first coordination shell. This choice of termination has been previously shown to result in ionization potentials and electron affinities that compare well with the measured ones for a range of semiconductor materials \cite{stevanovic_APL:2014, stevanovic_PCCP:2014}. As shown on the left of Fig.~\ref{fig:workflow}, the terminations do not necessarily have to be planar and can be corrugated (jagged blue line). In order to simulate surfaces (and interfaces) we created slabs of several layers of gallium oxide that are periodic parallel to the specified plane, $(\overline{2}01)$ or (100), but finite in the direction normal to that plane. Each slab is separated by 20\r{A} of vacuum such that there is no interaction between them. Since the slabs should ideally be semi-infinite, they are made to be as thick as computationally affordable. Once the surfaces were chosen, they were relaxed using density functional theory (DFT). The relaxed $\beta$-Ga$_2$O$_3$ structure was then fed into the structure matching algorithm described hereafter. 
 
On the platinum side (right of Fig.~\ref{fig:workflow}), there was always only one possible termination because there is a single atom in the FCC unit cell. We did not relax the surfaces because, in experiment, the metal is deposited directly onto the gallium oxide without ever forming a surface. We fed the algorithm every possible, symmetry inequivalent orientation of platinum within the Miller indices limits $-3 \leq h,k,l \leq 3$. For each of these orientations, we computed the ``goodness of match'' (described below), which then provided the best matched interface structures.

Finally, once the best Pt/Ga$_2$O$_3$ matching structures were found for $(\overline{2}01)$ and (100), we passivated the outside surface of gallium oxide using partial H atoms such that there was no surface states or dipole at the outside surface. We relaxed the full structure except for the layer of oxygen that is furthest from the interface using DFT. In all DFT calculations we used the Generalized gradient approximation (GGA) with the Perdew, Burke, Ernzerhof functional (PBE) and the projector augmented wave (PAW) pseudopotentials as implemented in the VASP software \cite{perdew1996generalized, blochl1994projector, kresse1996efficiency}. We calculated electronic properties with Heyd, Scuseria and Ernzerhof (HSE06) functional \cite{krukau2006influence} to obtain a more realistic description of the electronic structure including the bandgaps and to compute the Schottky barrier height.

\subsection{Matching Crystal Structures}
The structure matching algorithm, detailed in Ref.~\cite{therrien2020matching}, minimizes a given cost function (e.g., total displacement of all atom, energy,\dots) with respect to the atom-to-atom mapping and relative orientation of two sets of atoms. It has been shown to successfully find optimal transformation pathways between two materials \cite{therrien2020minimization}. In Ref.~\cite{therrien2020matching}, we demonstrated that the structure matching could also be used to find the orientation relationship and bonding pattern at two semi-coherent hetero-interfaces:  SiC(110)/Si(001) and yttria-stabilized zirconia YSZ(111) / Ni (111). In the current work, we used an improved version of this algorithm to find the best matching and orientation relationships between the termination planes of platinum and $\beta$-Ga$_2$O$_3$.

To find the optimal structure, the algorithm goes through each combination of termination planes as illustrated in the center of Fig.~\ref{fig:workflow}: $\beta$-Ga$_2$O$_3$$(\overline{2}01)$/Pt(100), $\beta$-Ga$_2$O$_3$$(\overline{2}01)$/Pt(110), $\beta$-Ga$_2$O$_3$ $(\overline{2}01)$/Pt(111), etc. For each combination, the matching is performed in the following way. First, two large spherical (in 2D circular) sections of the two crystals with $\sim$300-450 atoms are created. One set contains only the oxygen atoms that are at the surface of gallium oxide and the other one contains only the surface platinum atoms. In both cases, only atoms that are within 1\r{A} from the surface are considered. These sets are then matched directly without using any information related to the periodicity of the two crystals. Each oxygen atom is mapped to a platinum atom in an optimal manner using the Kuhn-Munkres algorithm \cite{kuhn1955}. The quantity that is being optimized is a two-body Lennard-Jones potential with an equilibrium radius 2.5 \r{A}. In an iterative process, the total Lennard-Jones energy (the ``goodness of matching'' function) is simultaneously minimized with respect to the relative orientation (angle), in-plane strain and atom-to-atom mapping (choice of a ``bonding'' pattern) between the two surfaces. 

The strains and rotation angle are represented by a $2{\times}2$ transformation matrix. In this case we limited the amount of strain to 8\% in each in-plane direction (principal strain), it can be either tensile or compressive. We  applied the strain fully on the platinum side of the interface leaving $\beta$-Ga$_2$O$_3$ intact. In the final step, once the optimal match is found the periodicity of the mapping is retrieved i.e., the two-dimensional unit cell (represented by a $2{\times}2$ matrix $C_{Ga_2O_3}$) that is common two both structures in their optimal orientation is found. This step can have the effect of increasing the strain above the threshold (8\%) such that the transformation matrix is exactly commensurate with both unit cells: $C_{Ga_2O_3} = TC_{Pt}$ where $C_{Ga_2O_3}$ and $C_{Pt}$ are supercells of the $\beta$-Ga$_2$O$_3$ and Pt in-plane unit cell respectively. For each combination, an optimal total bonding energy is obtained, its absolute value has no physical meaning, but it can be used to compare the quality of the match. As noted already, this procedure has been shown to produce realistic interface structures. The effect of strain will be discussed later in the text. The structure matching algorithm used here is available in the form of an open-source software called \textsc{p2ptrans} \footnote{\href{https://github.com/ftherrien/p2ptrans}{github.com/ftherrien/p2ptrans}}.      

\subsection{Computing the Schottky Barrier Height}
 In the past century, many models were developed to describe the physics of the Schottky barrier. First, Schottky \cite{schottky1939halbleitertheorie} and Mott \cite{mott1939theory} estimated that the SBH was equal to the difference between the work function of the metal and the electron affinity of the semiconductor (known as the Schottky-Mott rule). In other words, the bands are aligned such that the vacuum levels of the metal and the semiconductor coincide. Even though the simplicity of this model makes it a popular pedagogical tool to understand the formation of the barrier, it is unable to accurately describe most physical interfaces. Bardeen developed a more complex interface model that took into account the presence of surface states in the bandgap of the semiconductor \cite{bardeen1947surface}, which lead to the well-known Cowley-Cze equation \cite{cowley1965surface}. This theory explains how the existence of surface states may cause ``Fermi-level pinning'' but it relies on the existence of a physical gap between the semiconductors and the metal. In reality, the surfaces are close enough that surface states cannot remain unchanged (they become interface states) and the physical boundaries of the two materials are difficult to define. To account for this problem the metal induced gap states (MIGS) theory \cite{heine1965theory, tejedor1978simple, tersoff1984schottky} was proposed. It adapts Bardeen's theory to systems with intimate interfaces by replacing the gap between the materials with the length of the MIGS tails. However, there are doubts on the physical relevance of this adaptation \cite{tung2014physics}. Detailed accounts of Schottky Barrier theories can be found in Refs~\cite{tung2014physics, tung2001recent}. 
 
 More recently, SBH is computed using the first-principles electronic structure methods, typically density functional theory (DFT), by effectively measuring the difference in energy between the conduction band minimum in the semiconductor and the Fermi energy of the metal-semiconductor (MS) system at the actual (model) interface \cite{das1989electronic, fujitani1990schottky,lazzarino1998znse, berthod2003schottky, monch2004first, xu2020first}. Unlike the Schottky-Mott rule this method does take into account changes in the electronic structures of two materials due to interface formation. This is the method that we used in this paper. The approach implicitly makes the assumption that the band alignment at the interface, determined with DFT or in our case the hybrid HSE06 functional, remains completely unchanged by the doping of the semiconductor. 
 
Using the plane-wave basis in the electronic structure calculations poses two limitations. First, it requires a 2D periodic interface structure with relatively small number of atoms (couple of 100s). When there is no obvious lattice match and there is no known preferential orientation for the metal, it is difficult to choose an interface structure that is representative of the physical system. Because of that, it has remained challenging to assess the accuracy of \textit{ab initio} calculation at predicting the SBH. By using structure matching to create interface structures, we can not only justify our choice of a representative interface structure, but we can also relatively easily create interface structures in any orientation of Pt with varying numbers of atoms and measure their effect on the SBH. Such an analysis is presented further in this text. Second, the notorious bandgap underestimation of the standard approximations to DFT, namely the LDA and GGA, also make computations of the SBH difficult. However, it is presently possible to use more accurate and more computationally demanding electronic structure methods, such as the hybrid functions, on systems with multiple hundreds of atoms and overcame the limitations enforcing the use of LDA and GGA. In this way, a truly predictive method to compute SBH can be constructed, which is precisely what we are showcasing in this paper.

\section{Results}
\begin{figure}
\includegraphics[width=0.8\linewidth]{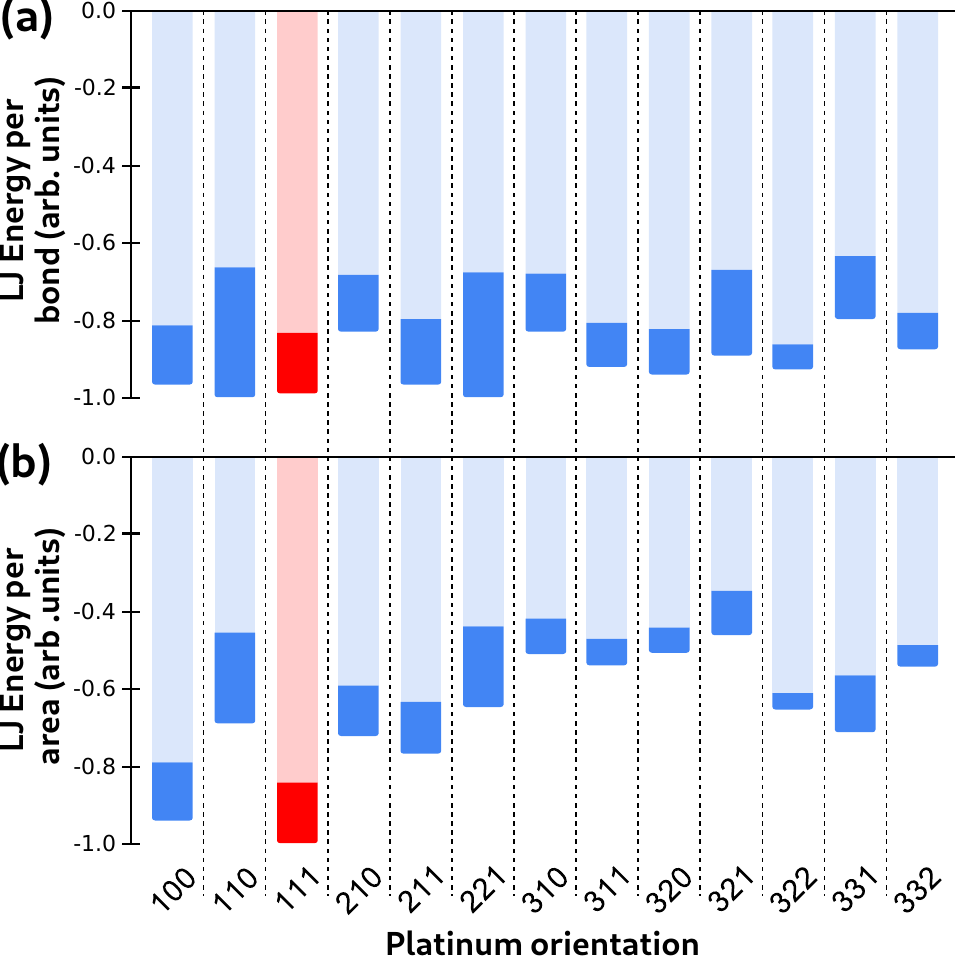}
\caption{\label{fig:distances}
Optimal Lennard-Jones energy (``goodness'' of the match) calculated between $\beta$-Ga$_2$O$_3$ $(\overline{2}01)$ and all, symmetry inequivalent Pt planes with Miller indices between -3 and 3 (a) per pair of matched Pt-O atoms (i.e. the Lennard-Jones ``bond''), and (b) per unit area of the interface. The values are normalized to 1.
}
\end{figure}
The average Lennard-Jones (LJ) energy or goodness of the match computed per the Pt-O pair of mapped atoms (i.e. the Lennard-Jones ``bond'') between $\beta$-Ga$_2$O$_3$ $(\overline{2}01)$ and the 13 orientations of platinum with Miller indices between -3 and 3 is shown in Fig.~\ref{fig:distances}(a). This is the output from the structure matching algorithm: lower the value, better the match. The full length of each bar represents the average goodness of the match per LJ bond when the Pt surface is strained. However, this does not account for the cost in energy from straining the Pt lattice to get the best matched interface. To quantify that effect, keeping the same bonds (mapping) at the interface, the strain in Pt is removed and the LJ energy is recomputed, i.e. instead of straining Pt, the bonds at the interface are stretched to compensate for the difference in unit cells between gallium oxide and platinum. Doing so provides an upper limit for the LJ energy where there is no contribution from strain (the pale blue part of each bar). In other words, the pale blue portion of each bar corresponds to a situation where the materials would be so stiff that no amount of strain would be present at the interface and the mismatch between the two structures would be fully accommodated by the bonds at the interface. The full length of the bars (dark blue + pale blue) on the other hand corresponds the situation where straining the atoms near the interface has no energetic cost up to the maximum allowed strain. The reality lies somewhere between the two (dark blue region). The wider the dark blue region the more strained the optimal result is for each orientation. 

\begin{figure}
\includegraphics[width=0.8\linewidth]{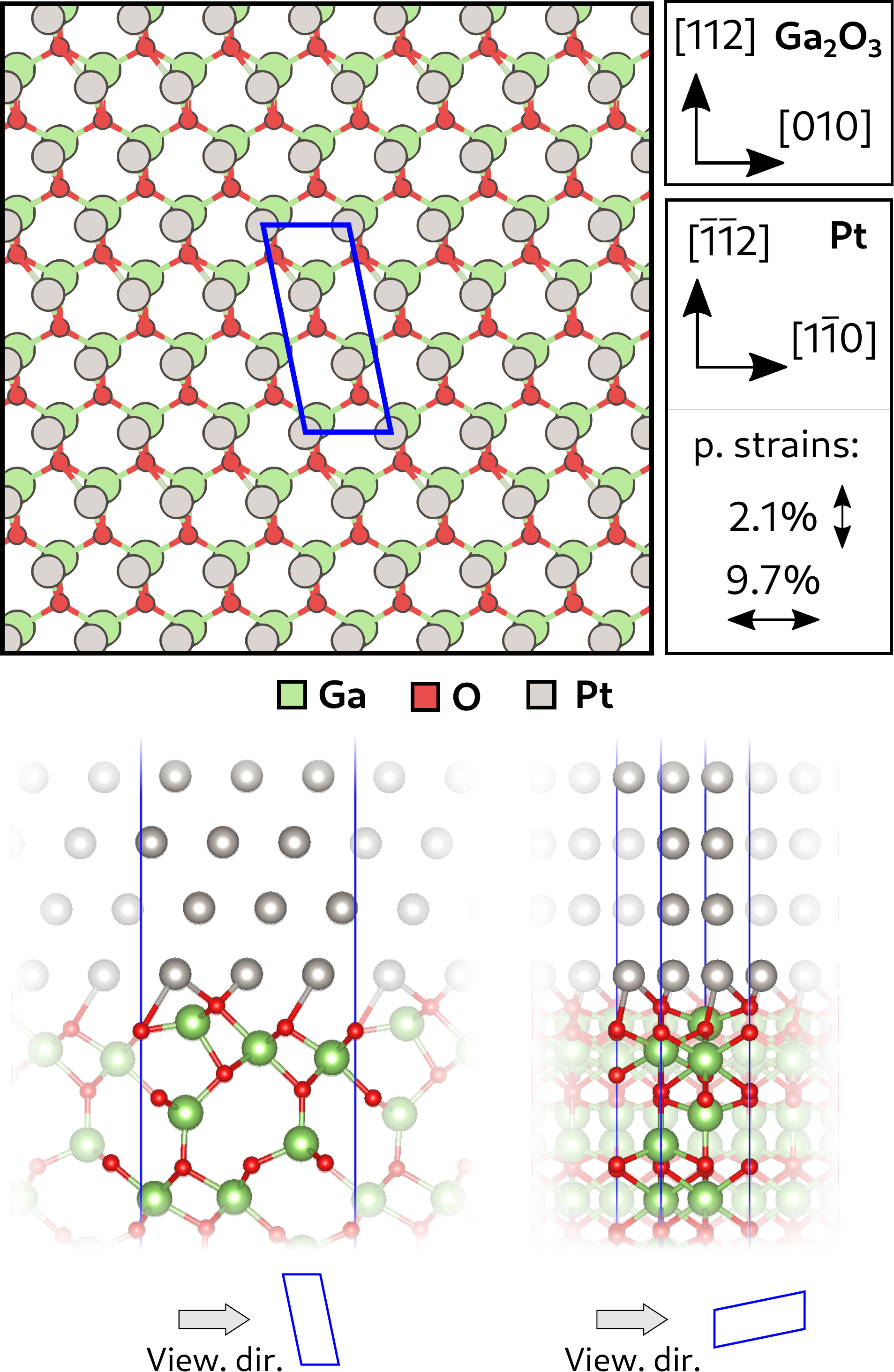}
\caption{\label{fig:result}
Optimally matched interface structure between $\beta$-Ga$_2$O$_3$ $(\overline{2}01)$ and Pt (111) corresponding to the red bars in Fig.~\ref{fig:distances}. The in-plane periodic unit cell of the interface is represented by a blue parallelogram. Outward pointing arrows indicate tensile strain. Strains apply to Pt only.  
}
\end{figure}

Looking at Fig.~\ref{fig:distances}(a), the Pt orientations that best match $\beta$-Ga$_2$O$_3$ $(\overline{2}01)$ in terms of the Pt-O mapping are: (110), (221), (111) and (100), with (110) and (221) needing large strains to accommodate a near perfect matching. Since the density of surface atoms is different for different orientations, it is more relevant to look at the LJ energy per unit of area [Fig.~\ref{fig:distances}(b)] since it represents a proxy for the interface energy. When doing so, it becomes clear that the (111) orientation of Pt is the best matching orientation, followed by (100). The optimal interface structure used in the rest of this paper, $\beta$-Ga$_2$O$_3$ $(\overline{2}01)$/Pt(111), is depicted in Fig.~\ref{fig:result}. As explained further in this section, the relatively high tolerance on strain allows for a smaller interface unit cell (shown in blue) where the hexagonal shape of the Pt (111) plane is distorted to match perfectly that of the $\beta$-Ga$_2$O$_3$ $(\overline{2}01)$ surface. Note that the interface in Fig.~\ref{fig:result} is an atomistic model/representation of the situation at flat $\beta$-Ga$_2$O$_3$ $(\overline{2}01)$$\beta$-Ga$_2$O$_3$ $(\overline{2}01)$ terraces. A more realistic representation of the microstructure including steps, kinks, etc. would require a large interface structure that would hardly be tractable with ab initio methods. 

\begin{figure*}
\includegraphics[width=0.8\linewidth]{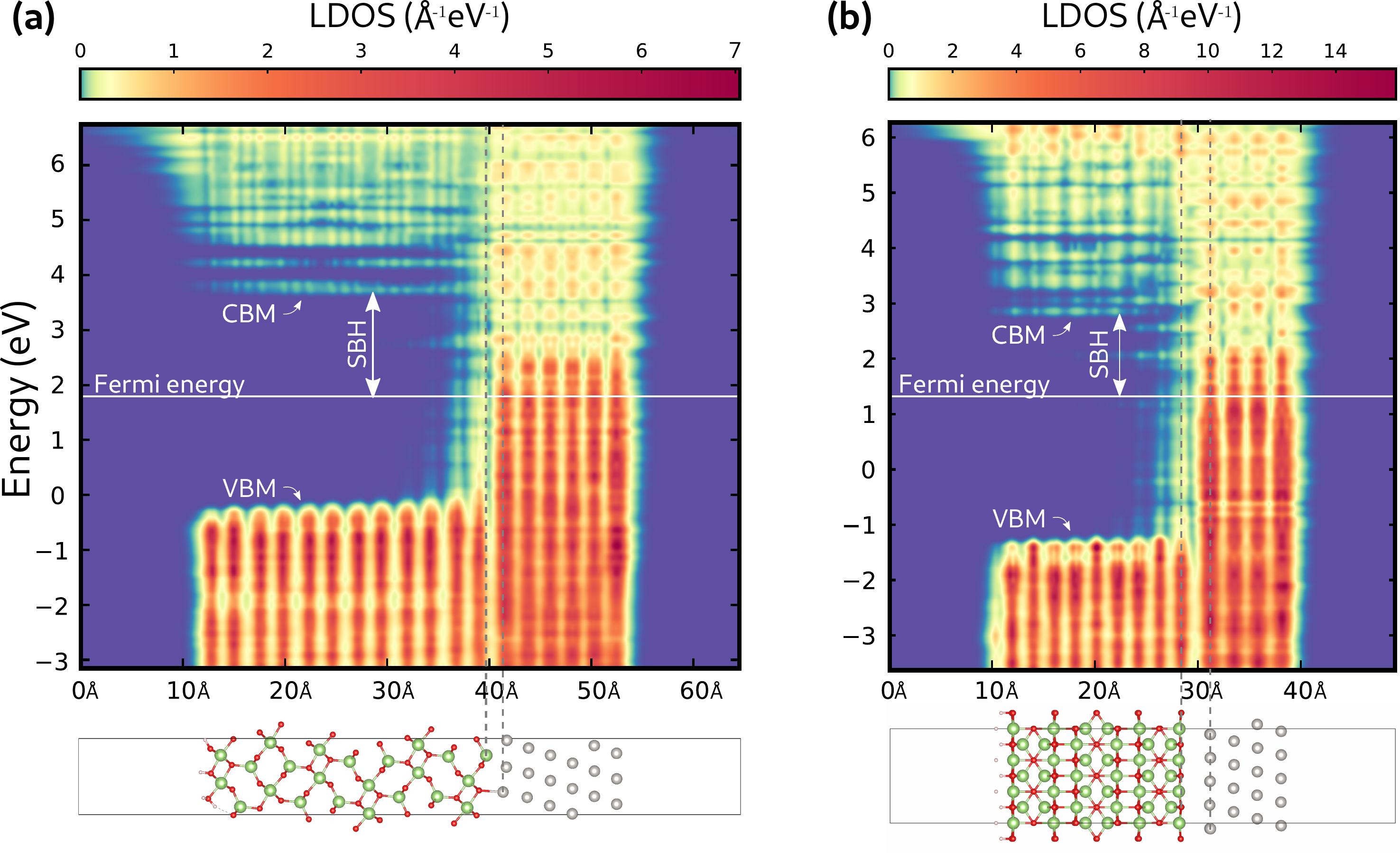}
\caption{\label{fig:ldos_1}
Local density of states for (a) the $\beta$-Ga$_2$O$_3$ $(\overline{2}01)$ and Pt (111) interface and (b) the $\beta$-Ga$_2$O$_3$ $(100)$ and Pt (111) interface. The Fermi energy is shown as a white line, the x-axis is in line with the corresponding relaxed structure.}
\end{figure*}

After relaxation of the interface structure, all the Pt atoms moved slightly away from gallium oxide as the Ga atom that was closest to the interface moved towards platinum retrieving its original bulk position [see the bottom of Fig.~\ref{fig:ldos_1}(a)]. From this relaxed structure, we obtained the position-dependent local density of states (LDOS) by computing the charge density for 100 energy intervals between $E_f$-5~eV and $E_f$+5~eV where $E_f$ is the Fermi energy. For each interval we integrated the charge density in-plane to obtain the density of states as a function z along the interface structure. The LDOS for the relaxed interface structure between  $\beta$-Ga$_2$O$_3$ $(\overline{2}01)$ and Pt (111) along with the relaxed crystal structure are presented in Fig.~\ref{fig:ldos_1}(a). 
The bandgap in gallium oxide is approximately 4~eV which is an underestimation compared to the experimental value (4.8~eV). This difference is due to the remaining underestimation of HSE for DFT relaxed $\beta$-Ga$_2$O$_3$. See the Supplemental Material (SM) \footnote{See Supplemental Material at [URL] for a complete discussion on bandgaps and finite size effects including Refs~\cite{cipriano2020quantum, yoffe1993low, kayanuma1988quantum} and for the local density of state and crystal structures for the following interfaces: $\beta$-Ga$_2$O$_3$ (100) / H.OH / Pt (111), $\beta$-Ga$_2$O$_3$ ($\overline{2}$01) / Pt (111) with reduced strain, $\beta$-Ga$_2$O$_3$ ($\overline{2}$01) / Pt (221) and $\beta$-Ga$_2$O$_3$ ($\overline{2}$01) / Pt (321). We also included all the structure files (POSCAR) used to calculate the local densities of state presented in this paper.} for a more detailed discussion on the bandgap in slabs of finite sizes.

In this plot, we can directly measure the Schottky barrier height from the Fermi energy to the conduction band minimum on the gallium oxide side, by assuming that the conduction band minimum is the first visible state (pale blue) that spans the entire gallium oxide structure. We predict Schottky barrier height of approximately 1.9~eV. Since the bandgap is somewhat underestimated, we can expect that the true SBH would be slightly higher, an effect that would be compensated in part by the image charge effect which is of the order of 0.1-0.2~eV \cite{oishi2015high, sze1964photoelectric}, depending on the level of doping and the applied voltage. Even with the underestimated bandgap, this is much larger than the reported experimental SBH for that interface: 1.05~eV (from current-voltage characteristics) \cite{yao2017electrical, fu2018comparative}. We also obtained the position of the conduction band minimum (CBM) and valence band maximum (VBM) by shifting the bulk values by an amount corresponding to the average electrostatic potential in gallium oxide and obtained the same SBH (see SM \cite{Note2}). Here, as others have done before \cite{das1989electronic, fujitani1990schottky,lazzarino1998znse, berthod2003schottky, monch2004first, xu2020first}, we make the assumption that position of the CBM with respect to the metal energy levels at the interface will not change with doping. We do not expect to see any appreciable band bending at this scale since the depletion region is much larger than the structures used here. If we were able to see it, the bands of doped gallium oxide would bend down such that the Fermi level would be close to the CBM far away from the interface.

In order to validate our method, we applied the same workflow (Fig.~\ref{fig:workflow}) to the (100) orientation of $\beta$-Ga$_2$O$_3$ which has been studied both experimentally \cite{he2017schottky} and computationally \cite{xu2020first}. From structure matching we obtained Pt (111) as the optimal match once again. It is the same Pt orientation that was used in Ref~\cite{xu2020first} where it was chosen based on the fact that it is commonly used in experiment. The Ga$_2$O$_3$ (100) being a very stable surface, relaxation of the surface and the interface structure do not significantly change the atomic layout. The local density of state of the relaxed interface structure is presented in Fig.~\ref{fig:ldos_1}(b). One may notice that the bandgap in the (100) orientation is slightly larger than in the $(\overline{2}01)$ orientation. This can be explained by spurious finite size effects that are more important in the (100) orientation which is thinner despite requiring more atoms (see SM \cite{Note2} for more details). In this case the Schottky barrier height is significantly lower: 1.5~eV. This value is much closer to experimental measurements in that orientation: 1.4~eV (from current-voltage characteristics) \cite{he2017schottky} and the theoretical estimate of 1.4~eV obtained by Xu et al. with nearly identical parameters. This provides confidence in our ability to predict the Schottky barrier height but it begs the question: ``Why is the predicted Schottky Barrier height in the $(\overline{2}01)$ so much higher than the one measured in experiment?''  
\subsection{Adsorption of Water Products}
\begin{figure}
\includegraphics[width=0.85\linewidth]{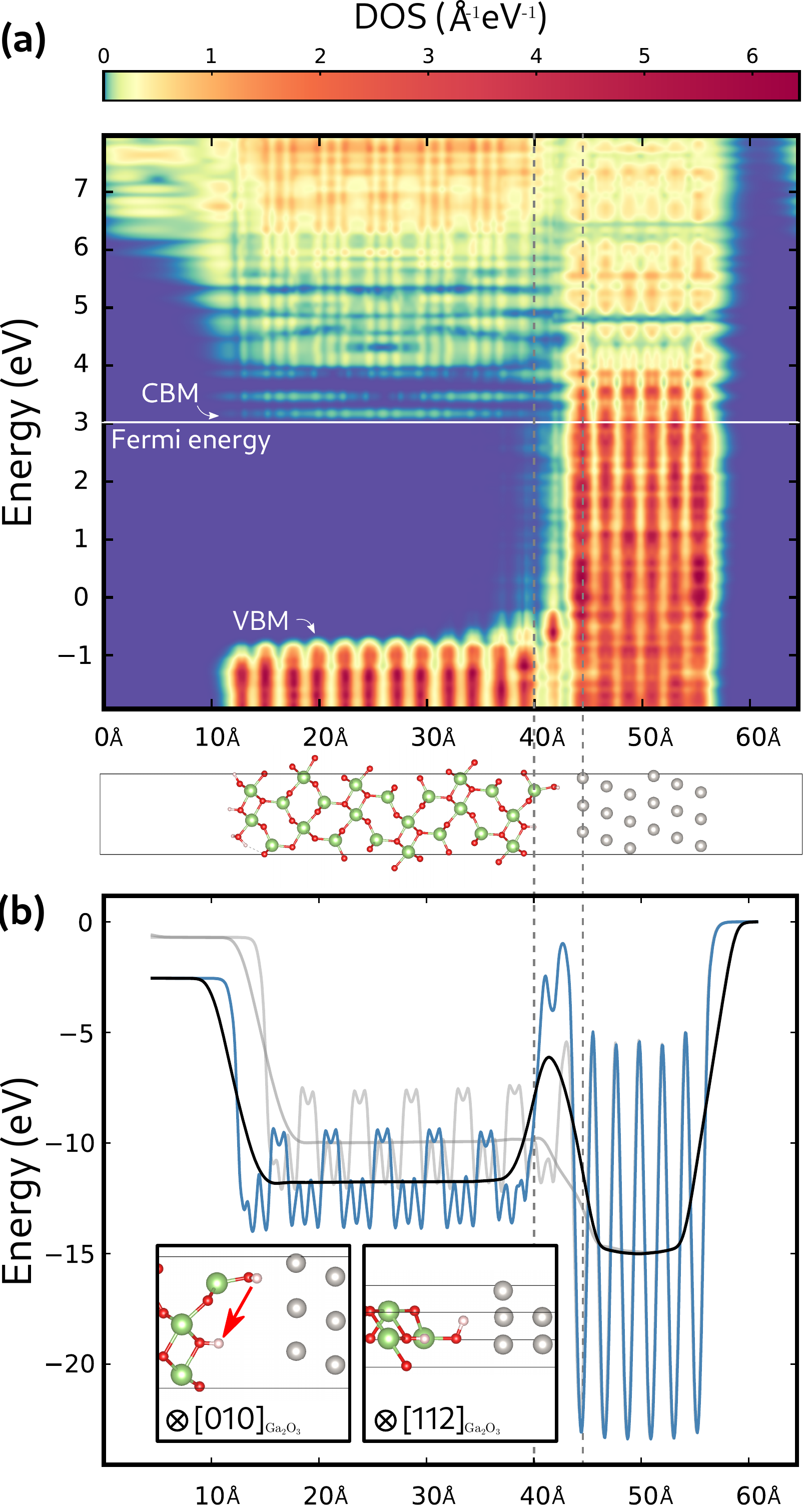}
\caption{\label{fig:ldos_2}
Electronic properties of a $\beta$-Ga$_2$O$_3$ $(\overline{2}01)$/H.OH/Pt(111) interface. (a) Local density of states showing the Fermi energy as a white line. (b) Electrostatic potential in blue with its average in black. The pale gray lines are the electrostatic potential and average electrostatic potential for the interface shown Fig.~\ref{fig:ldos_1}(a). The inset in panel (b) shows a zoomed view of the interface from two directions. The red arrow shows the direction of the dipole between OH and H (physics convention).}
\end{figure}
To answer this question, we looked at possible contaminants that could have been adsorbed at the surface during preparation and have an influence on the SBH. Since $\beta$-Ga$_2$O$_3$ samples are usually washed with deionized water before platinum is deposited \cite{tellekamp2020growth, heinselman2021performance} we focused on water-like products. Moreover, commercially available $\beta$-Ga$_2$O$_3$ $(\overline{2}01)$ substrates are known to be H and OH terminated \cite{gazoni2020relationship}. Therefore, to model this situation, we created a structure where the two broken bonds per unit cell at the Ga$_2$O$_3$ surface were connected to H$^+$ and OH$^-$ groups forming decomposed water (H.OH). The actual surface may contain varying proportions of H and OH depending on the treatment, most of them coming in equal parts from the decomposition of water molecules \cite{anvari2018density, ma2017monoclinic, gazoni2020relationship}. Since we focus on the typical deionized water treatment, the proportion of 50\% is appropriate. It also has the important advantage of being charge-neutral.

Upon relaxation, the decomposed water remained stable at the surface. We then added the Pt layers in the same orientation as in Fig.~\ref{fig:ldos_1}(a). Once again the decomposed water remained stable upon relaxing the interface. Note that, after relaxation, the attached OH$^-$ at the interface is almost parallel to the surface [see the inset of Fig.~\ref{fig:ldos_2}(b)] which is consistent with previous work \cite{anvari2018density}. 

The local density of states (LDOS) of that interface is shown in Fig.~\ref{fig:ldos_2}(a). It is immediately evident that the Schottky barrier height is reduced to practically zero by the presence of the adsorbed decomposed water; the Fermi energy sits directly under the conduction band minimum. In Fig.~\ref{fig:ldos_2}(b) the electrostatic potential of the $\beta$-Ga$_2$O$_3$ $(\overline{2}01)$/H.OH/Pt(111) interface, in blue, and the electrostatic potential of the $\beta$-Ga$_2$O$_3$ $(\overline{2}01)$/Pt(111) interface, in gray, are overlayed using the vacuum level on the Pt side as a reference. One can see that the entire electrostatic potential on the gallium side of the interface is shifted by an amount corresponding to the SBH in Fig.~\ref{fig:ldos_1}(a). The average potential on both sides of the interface remains flat which means the additional dipole that causes the shift in potential is located at the interface. This added dipole illustrated in the inset of Fig.~\ref{fig:ldos_2} goes from the negatively charged OH- to the positively charged H+ on the corrugated Ga$_2$O$_3$ surface.

In Fig.~\ref{fig:ldos_2} the concentration of adsorbed H.OH is very high; it assumes every single interface cell (Fig.~\ref{fig:result}) adsorbs one water molecule. Therefore, certain properties of the interface structure in Fig.~\ref{fig:ldos_2} are not representative of the real interface where the concentration of adsorbed water is much lower. For example, there is a large physical gap between the Pt atoms and the Ga$_2$O$_3$ atoms in Fig.~\ref{fig:ldos_2}; in reality, this gap would be present only close to the OH group and Pt atoms far away from it would be closer to Ga$_2$O$_3$ as in Fig.~\ref{fig:ldos_1}(a). In the same way, the measured SBH of the full interface would likely fall between that of Fig.~\ref{fig:ldos_1}(a) and that Fig.~\ref{fig:ldos_2}. One way to estimate the effective Schottky barrier is to compute the shift in electrostatic potential due to a finite concentration of dipoles at the interface. For example, using the position of the added H and O atoms and by solving the Poisson equation in 1D, we can estimate the charge density necessary to shift the dipole by 0.85~eV (to go from the ideal 1.9~eV to the experimental 1.05~eV barrier). We find that there would need to be about one adsorbed H.OH every 14 unit cells ($5.3\times10^{-11}$~mol/cm$^2$) to obtain a SBH of 1.05~eV. This offers a simple understanding of the concept of effective SBH in the presence of surface dipoles. 

Another interpretation is that the barrier itself is inhomogeneous consisting of regions with high SBH corresponding to the interface in Fig.~\ref{fig:ldos_1}(a) and smaller regions with low SBH corresponding to the interface in Fig.~\ref{fig:ldos_2} (or other interface defects). The effective or apparent SBH of the device, measured in experiment, would lie between the two values depending on the concentration of adsorbed water and the experimental method used to measure the SBH \cite{werner1991barrier, tung1992electron}. In fact, regions of the surface (``patches'') with low SBH have strong effect on the observed current-voltage behavior of the device and consequently the measured value of the barrier height \cite{tung1992electron}. 

There are two ways to experimentally measure the Schottky barrier height: using the capacitance-voltage (C-V) characteristics or using the current-voltage (I-V) characteristics. Measurements using the C-V characteristic are less sensitive to fluctuations and closer to the average SBH whereas I-V-based measurements are more representative of the effective barrier for current flow \cite{tung1991electron, yao2017electrical}. For example, Ref~\cite{yao2017electrical} reported a barrier height of 1.05~eV using the I-V characteristic and a barrier height of 1.9~eV using the C-V characteristic. This is consistent with our result since the concentration of adsorbed H.OH is low, the average SBH measured with C-V is near the calculated ideal value of 1.9~eV, whereas the effective value measured with I-V is strongly reduced by the presence of a few near zero patches. Note that the exactitude of the agreement of the C-V measurement with our predicted value (1.9~eV in both cases) may depend on small differences in experimental preparation since a more recent study \cite{fu2018comparative} which also measured 1.05~eV with the I-V characteristic found a 1.20~eV SBH with the C-V characteristic.

As mentioned before the (100) surface is more stable than the $(\overline{2}01)$ surface. We computed their energies to be about 0.03~eV/\r{A}$^2$ and 0.06~eV/\r{A}$^2$ respectively, in agreement with other theoretical data \cite{anvari2018density}. This difference in surface energy has an influence on their respective growth rate \cite{mazzolini2020substrate} and their ability to adsorb water products during surface preparation. In fact, the adsorption of H.OH on the (100) surface is unstable with respect to the adsorption of water \cite{ma2017monoclinic, zhou2016exploring} which could explain why our estimate of the SBH in the (100) orientation is more accurate: there is simply not as many H.OH on the surface to reduce the effective barrier height. Moreover, we created a $\beta$-Ga$_2$O$_3$ (100) surface with H.OH by connecting the 6 broken bonds per unit cell with H$^+$ and OH$^-$ (3 with H$^+$ and 3 with OH$^-$) as we did for the (111) surface (which had 2 broken bonds). The structure was stable after initial relaxation, but one of the three H.OH recombined to form a water molecule upon adding the Pt layers. We computed the LDOS of that interface and found that the addition of H.OH (2 H.OH and 1 H$_2$O after relaxation) had no significant impact on the barrier height which further demonstrates that the presence of water-like products on that surface has a less significant impact on the SBH. 

The fact that H$^+$ and OH$^-$ at the $(\overline{2}01)$ $\beta$-Ga$_2$O$_3$ surface have a strong influence on the SBH could explain why surface treatment with acids have been shown to increase the measured barrier height \cite{yao2017electrical}. This means that the apparent barrier height of an acid-cleaned interface with less OH would be higher because of the reduction in the number of patches with near zero effective barrier heights. Our results suggest that surface preparation should aim at reducing as much as possible the presence of decomposed water on the surface which offers a route to obtaining a high Schottky barrier without adding any additional dipole layers (e.g. PdCoO$_2$\cite{harada_APLM:2020}). Methods that avoid the use water completely should be considered.

Other preparation methods such as annealing in oxidizing condition and oxygen plasma treatment are known to have an important impact on the rectifying properties of metal/oxide interfaces. These methods contribute to removing oxygen vacancies near the oxide surface which has been shown to influence the Schottky/ohmic behavior of other metal oxide devices \cite{mosbacker2005role, allen2008influence, schafranek2008barrier, schultz2018influence}. Yet, they may also reduce water products adsorbates at the surface. For example, it was found that most of the measured increase in SBH that results from oxygen plasma treatment of ZnO in a ZnO/Au interface is attributable to the removal of surface adsorbates; the reduction of oxygen vacancies having only a small impact \cite{mosbacker2005role}. This means that some effects associated with the reduction of oxygen vacancies may in fact be due to the phenomena discussed in this paper. In the same way, it is possible that oxygen vacancies in Ga$_2$O$_3$ play a role in explaining the observed discrepancy between the predicated and measured SBH discussed here. In fact, studies have found that oxygen vacancies can change the Schottky behavior of other Ga$_2$O$_3$/metal systems \cite{guo2014oxygen, lingaparthi2019surface, lingaparthi2020effects}.

However, we believe that the presence of water adsorbates discussed in this section is the dominant effect since the removal of oxygen vacancies has a relatively small impact on the SBH itself \cite{lingaparthi2020effects}. The change in rectifying properties is mainly due to the change in tunneling width \cite{mosbacker2005role, guo2014oxygen, lingaparthi2020effects} (or Schottky barrier width) which is an important interface property that is independent from the SBH. Plus, the presence of oxygen vacancies could not easily explain the differences between the predicate (100) and ($\overline{2}01$) barrier heights. Moreover, it has been shown experimentally that the presence of H and OH at the ($\overline{2}01$) Ga$_2$O$_3$ surface changes the band bending near the surface by as much as 1~eV \cite{swallow2019transition, gazoni2020relationship} which is of the same magnitude as the effect measured here. Studying the underlying mechanisms through which oxygen vacancies change rectifying properties of our system would be an interesting research avenue. In any case, an approach such as oxygen plasma treatment should be considered to improve the general rectifying properties of the interface as it has a positive effect on both phenomena i.e., it removes adsorbates while creating oxidizing conditions to fill oxygen vacancies.

\subsection{Effect of Strain and Platinum Orientation} \label{sec:PtOR}

\begin{figure}
\includegraphics[width=0.8\linewidth]{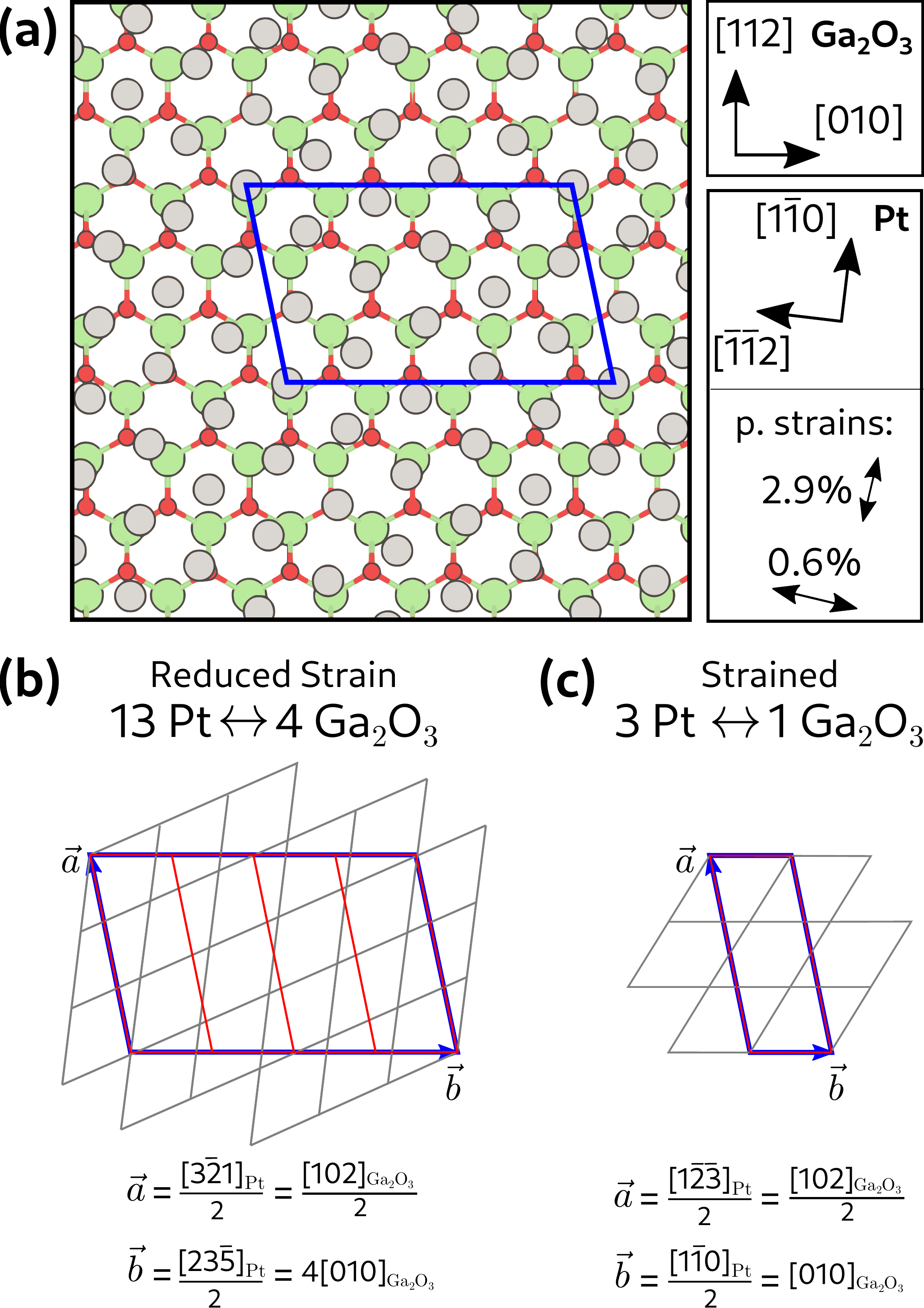}
\caption{\label{fig:strict}
Low and High strain $\beta$-Ga$_2$O$_3$ $(\overline{2}01)$ and Pt (111) interfaces. 
Panel A shows a top view of the best matching interface with a lower strain (stricter constraint). Panel (b) and (c) show the orientation of the in-plane unit cell in terms of the Ga$_2$O$_3$ and Pt coordinates for the reduced strain case (b) and the case with higher strain used throughout this paper (c). The in-plane unit cells of the interface structure, gallium oxide and Pt are shown in blue, red and gray respectively. Outward pointing arrows indicate tensile strain. Strains apply to Pt only.}
\end{figure}

To show the effect of the choice of the maximal allowable strain when matching interfaces and to provide confidence in our result, we compared best matching interfaces with different strain tolerances. Let us focus on two structures between $\beta$-Ga$_2$O$_3$ $(\overline{2}01)$ and Pt (111): one with a strain of less than 8\%, used throughout this paper and presented in Fig.\ref{fig:result}, and one with a strain of less than 2.5\% presented in Fig.~\ref{fig:strict}(a). In both cases, the in-plane strain is fully applied to the platinum side of the interface; gallium oxide retains its DFT-relaxed in-plane lattice parameters.   

We calculated the SBH for the interface with less than 2.5\% strain (panel B) and found a value of 2.4~eV. Although it is a slightly higher value than 1.9~eV for 8\% strain, it is consistent with the over-estimation of the SBH and it can be explained in part by the more important finite size effects caused by the thinner Ga$_2$O$_3$ which is required to complete the calculation with this larger cell.

Notice that one of the principal strains in Fig.~\ref{fig:result} is higher than 8\% and one of the principal strains in Fig.~\ref{fig:strict} is higher than 2.5\%. This is because the final interface unit cell must be a supercell of both the unit cells of gallium oxide and platinum as illustrated in panel B and C of Fig.~\ref{fig:strict}. This condition is true only for specific principal strains where vertices of the strained Pt and Ga$_2$O$_3$ coincide; Fig.~\ref{fig:result} the principal strains of 9.7\% and 2.1\% are the closest ones to the optimized strains--which are necessarily lower than 8\%--that fulfill this condition.

The advantage of allowing for a higher strain is that the resulting interface structures are smaller which is desirable for ab initio calculations. In Fig.~\ref{fig:result} and Fig.~\ref{fig:strict}(c), the interface cell, in blue, is comprised of only one $\beta$-Ga$_2$O$_3$ $(\overline{2}01)$ cell. In comparison, the optimal interface unit cell with strain illustrated in Fig.~\ref{fig:strict} (a) and (b) is composed of 4 $\beta$-Ga$_2$O$_3$ $(\overline{2}01)$ unit cells. In that structure, the Pt is rotated by about 90\degree anticlockwise. The strain is 2.9\% in one direction and close to zero in the other. As the strain tolerance decreases, the ratio of Pt to Ga$_2$O$_3$ changes. The high strain structure [Fig.~\ref{fig:strict}(c)] has a ratio of 1:3 = 33.33\% whereas the lower strain structure [Fig.~\ref{fig:strict}(b)] has a ratio of 4:13 = 30.77\%. At even lower strains this ratio would tend to 30.03\%, the exact ratio of specific areas between Pt (111) and $\beta$-Ga$_2$O$_3$ $(\overline{2}01)$. In the same way approximating an irrational number with an integer fraction requires larger integers for a higher precision, the size of the cell increases as the allowable strain decreases. 

Finally, we evaluated the SBH for two other orientations in Pt: (221) and (321). We chose (221) because it has excellent bonding energy per bond [Fig.~\ref{fig:distances}(a)] and a relatively good average energy per area [Fig.~\ref{fig:distances}(b)]. It also has the particularity of having a saw-tooth corrugated surface that is perfectly matched with the corrugated surface of gallium oxide. We chose the (321) surface because it is the worst matching interface so to see the effect of a ``bad'' match on the barrier height. For the (221) surface, we obtained a SBH of 1.9~eV and for the (321) interface, we obtained a SBH of 2.1~eV. These values do not differ significantly from the 1.9~eV barrier calculated for the Pt(111) orientation.

The test cases presented in this section give us better confidence that our main findings are applicable to real systems where Pt has no preferential orientation and where the strain at the interface might vary between different grains. Local density of states plots and their corresponding structures for all three interfaces presented in this section are available in the Supplemental Material \cite{Note2}.   

\section{Conclusion}

In this work, we created a plausible interface structure between face centered cubic platinum and $\beta$-Ga$_2$O$_3$ $(\overline{2}01)$ using structure matching \cite{therrien2020matching}. We found that the best matching orientation of Pt is (111). We then computed the local density of states (LDOS) of the newly created interface structure using density functional theory and predicted a Schottky barrier height (SBH) of 1.9~eV which is much higher than the experimentally measured value of 1.05~eV (current-voltage characteristics) \cite{yao2017electrical, fu2018comparative}. To validate our methodology, we applied the same process to the (100) orientation of $\beta$-Ga$_2$O$_3$, which has been studied computationally \cite{xu2020first} and experimentally \cite{he2017schottky}, and found excellent agreement. Why is the predicted SBH so overestimated in $(\overline{2}01)$ when it is relatively accurate in (100)? To answer this question, we created a $\beta$-Ga$_2$O$_3$ $(\overline{2}01)$/H.OH/Pt(111) interface and found that it had a zero barrier height (ohmic behavior). The presence of decomposed water in some regions of the interface from conventional surface cleaning methods could have the effect of reducing the effective SBH of the device, which would explain the discrepancy between predicted and measured values of the SBH in the $(\overline{2}01)$. 

Our results are consistent with the fact that pretreatment of the $\beta$-Ga$_2$O$_3$ $(\overline{2}01)$ surface with acids has a positive effect on the barrier height. They also confirm, as pointed out in Ref~\cite{yao2017electrical} the importance of surface preparation during the fabrication of $\beta$-Ga$_2$O$_3$ devices. In fact, our results indicate that a perfectly clean $\beta$-Ga$_2$O$_3$ $(\overline{2}01)$/Pt interface could almost double the SBH ($\approx$2~eV). Since the barrier height of a metal semiconductor interface is the property that has the most impact on its electrical characteristics \cite{tung2014physics}, this could help design and fabricate high power and high temperature electronic devices with lower leakage current that could sustain larger breakdown fields.

Finally, we varied the strain and the orientation of Pt and found that it has little impact of the predicted Schottky barrier height. This shows how our approach using structure matching and the local density of states is useful in easily creating various interface structures and testing the influence of different physical parameters on the system. It offers a fully predictive and convenient way to study electronic properties that can be applied to any solid-solid hetero-interface.

\begin{acknowledgments}

This work was authored in part at the National Renewable Energy Laboratory (NREL), operated by Alliance for Sustainable Energy, LLC, for the U.S. Department of Energy (DOE) under Contract No. DE-AC36-08GO28308. Funding provided by the Office of Energy Efficiency and Renewable Energy (EERE) Advanced Manufacturing Office (AMO). Method development was supported by the National Science Foundation Grant No. DMR-1945010. The research was performed using computational resources sponsored by the DOE’s Office of Energy Efficiency and Renewable Energy located at NREL and using the Colorado School of Mines high performance computing resources. The views expressed in the article do not necessarily represent the views of the DOE or the U.S. Government.

\end{acknowledgments}

\bibliography{reference}

\end{document}